\documentclass[showpacs,showkeys,aps,prd,superscriptaddress,nofootinbib]{revtex4}

\usepackage{graphicx}
\usepackage{amsmath}
\usepackage{epsfig}

\def\to           {\ensuremath{\rightarrow}}
\def\B            {\ensuremath{B}}
\def\Bz           {\ensuremath{B^{0}}}
\def\Bzb          {\ensuremath{\overline{B}^{0}}}
\def\piz          {\ensuremath{\pi^{0}}}
\def\Bp           {\ensuremath{B^+}}

\def\Btag         {\ensuremath{B_{\rm tag}}}
\def\Brec         {\ensuremath{B_{\rm rec}}}
\def\aone         {\ensuremath{a_1}}
\def\hh           {\ensuremath{h^+h^{\prime -}}}

\def\Bztorhoprhom {\ensuremath{\Bz \to \rho^+ \rho^- }}

\def\Bztorhozrhoz {\ensuremath{\Bz \to \rho^0 \rho^0 }}
\def\Bztorhoprhoz {\ensuremath{\B^\pm \to \rho^\pm \rho^0 }}

\def\Bztopippim   {\ensuremath{\Bz \to \pi^+ \pi^- }}
\def\Bztopizpiz   {\ensuremath{\Bz \to \piz \piz }}

\def\CP      {\ensuremath{ CP }}
\def\CPV     {\ensuremath{ CPV }}

\def\C       {\ensuremath{ C }}
\def\S       {\ensuremath{ S }}
\def\acp     {\ensuremath{ {\cal A}_{CP} }}
\def\clong   {\ensuremath{ C_{L} }}
\def\slong   {\ensuremath{ S_{L} }}
\def\ctran   {\ensuremath{ C_{T} }}
\def\stran   {\ensuremath{ S_{T} }}
\def\fL      {\ensuremath{ f_L }}
\def\ptrue   { \fL }
\def\mes     { \ensuremath { M_{ES} } }
\def\deltat  { \ensuremath{\Delta t }}
\def\deltaz  { \ensuremath{\Delta z }}
\def\deltae  { \ensuremath{\Delta E }}
\def\deltamd { \ensuremath{\Delta m_d }}
\def\penguin   {\ensuremath{ {(\rm penguin)} } }
\def\deltaalpha {\ensuremath{\delta\alpha_{\penguin}}}
\def\alphaeff{ \ensuremath {\alpha_{\mathrm{eff}}}}

\def\br        {\ensuremath{ {\cal {B}}}}
\def\ifb       {\ensuremath{ { fb^{-1} }}}
\def\iab       {\ensuremath{ { ab^{-1} }}}

\def\vckm        {\ensuremath{ {V}_{CKM}}}

\def\epem      {\ensuremath{ { e^+ e^- } } }
\def\qqbar     {\ensuremath{ { q \overline{q} } } }

\newcommand{\e}      [1]  {\ensuremath{\times10^{#1}}}
\newcommand{\su}     [1]  {\ensuremath{SU(#1)}}

\newcommand\vud {\ensuremath{V_{ud}}}
\newcommand\vus {\ensuremath{V_{us}}}
\newcommand\vub {\ensuremath{V_{ub}}}

\newcommand\vcb {\ensuremath{V_{cb}}}

\input babarsym

\newcommand{\SLACPubNumber} {11607}

\def\figurebox#1#2#3{%
    \def\arg{#3}%
    \ifx\arg\empty
    {\hfill\vbox{\hsize#2\hrule\hbox to #2{\vrule\hfill\vbox to #1{\hsize#2\vfill}\vrule}\hrule}\hfill}%
    \else
    {\hfill\epsfbox{#3}\hfill}%
    \fi}

\begin{document}


\begin{flushleft}
SLAC-PUB-\SLACPubNumber\\
\end{flushleft}

\title{Measurement of the CKM angle {\boldmath $\alpha$} with the B-factories.}

\author{\footnotesize Adrian Bevan}

\address{Physics Department, Queen Mary, University of London, Mile End Road, E1 4NS, UK.$\,^\dag$ \\
Physics Department, University of Liverpool, Oxford Street, Liverpool, L69 7ZE, UK.\\
$\,^\dag$ current address.\\
bevan@slac.stanford.edu}

\date{\today}

\begin{abstract}

\B-meson decays involving $b \to\ u$ transitions are sensitive to
the Unitarity Triangle angle $\alpha$ (or $\phi_2$).  The
\B-factories at SLAC and KEK have made significant progress toward
the measurement of $\alpha$ in recent years.  This paper
summarizes the results of the \B-factories' constraints on
$\alpha$.


\end{abstract}


\maketitle

\section{Introduction}

\CP\ violation (\CPV) was first seen in the decay of neutral
kaons~\cite{christenson}.  It was shown some time ago that \CPV\
is a necessary but insufficient constraint in order to generate a
net baryon anti-baryon asymmetry in the universe~\cite{sakharov}.
In the ensuing years there has been a tremendous amount of
activity by the high energy physics community to better understand
the role of \CPV\ in our model of nature $-$ the Standard Model of
Particle Physics (SM). 

\begin{figure}[!h]
\begin{center}
  \resizebox{8cm}{!}{\includegraphics{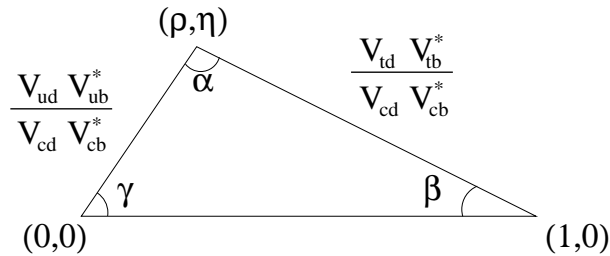}}
\end{center}
 \caption{The Unitarity Triangle.}
\label{fig:unitarity_triangle}
\end{figure}

\CPV\ in the SM is described by a single complex phase in a $3
\times 3$ quark-mixing matrix, \vckm, called the  CKM~\cite{CKM}
matrix:
\begin{eqnarray}
\vckm =
\left( \begin{array}{ccc}
 V_{ud} & V_{us} & V_{ub} \\
 V_{cd} & V_{cs} & V_{cb} \\
 V_{td} & V_{ts} & V_{tb}
\end{array}\right).
\end{eqnarray}
\vckm\ describes the coupling of the $u$, $c$ and $t$ quarks to
$d$, $s$ and $b$ quarks, which is mediated by the exchange of a
\W\ boson.  In \B-meson decays the interesting \CP\ violating
parameters of the SM are related to the angles ($\beta$, $\alpha$,
and $\gamma$)\footnote{The Belle Collaboration use the notation
$\phi_1$, $\phi_2$, and $\phi_3$ for the angles of the Unitarity
Triangle.} and sides of the so-called Unitarity Triangle (as shown in 
Figure~\ref{fig:unitarity_triangle}).  The angles are defined as:
\begin{eqnarray}
\alpha \equiv \arg\left[-V_{td}^{}V_{tb}^\star/V_{ud}^{}V_{ub}^\star\right], \\
\beta  \equiv \arg\left[ -V_{cd}V_{cb}^\star / V_{td}^{}V_{tb}^\star\right], \\
\gamma \equiv \arg\left[ -V_{ud}^{}V_{ub}^\star / V_{cd}V_{cb}^\star \right].
\end{eqnarray}

This review is a summary of the experimental constraints on the
Unitarity Triangle angle $\alpha$ obtained from \B-meson decays
involving $b \to u$ transitions. The CKM angle $\beta$ is well 
known and consistent with SM predictions~\cite{sin2beta}. Any 
constraint on $\alpha$ constitutes a test of the SM description 
of quark mixing and \CPV\ in \B-meson decays.  A significant 
deviation from SM expectation would be a clear indication of 
new physics. It is possible to obtain a SM prediction of $\alpha$
from indirect constraints by combining measurements of the CKM
matrix elements $|\vus|$, $|\vud|$, $|\vub|$, and $|\vcb|$, CPV\ in
mixing from neutral kaons, $\B$-$\Bbar$ mixing in $\B_d$ and
$\B_s$ mesons and the measurement of $\sin 2\beta$ from $b \to
c\overline{c}s$ decays. The SM predictions for $\alpha$ are $(98.2
\pm 7.7)^\circ$ from~\cite{utfitter}
 and $(97^{+13}_{-19})^\circ$~\cite{ckmfitter}.

\begin{figure}[!h]
\begin{center}
  \resizebox{10.0cm}{!}{\includegraphics{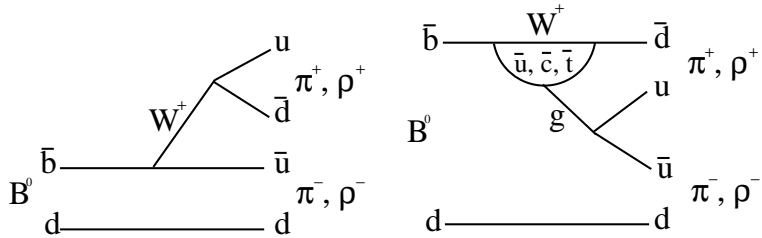}}
\end{center}
 \caption{The tree (left) and gluonic loop (right) contributions to $B \to \hh$ decays.}
\label{fig:feynman_diagrams}
\end{figure}

Measurements of $\alpha$ have recently been performed at the \B-factories, the \babar\
experiment~\cite{babar_nim} at SLAC and the Belle
experiment~\cite{belle_nim} at KEK. The \B-factories study the
decays of \B-mesons produced in the process $e^+e^- \to
\Upsilon(4\rm S) \to \B\overline{B}$, where $\B\overline{B}$ is either 
$\B^+\B^-$ or $\Bz\overline{B}^0$. The neutral \B-mesons are
produced in a correlated P-wave state. Until one of the \B-mesons
in an event decays, there is exactly one \Bz\ and one \Bzb\ meson.
The main goals of the \B-factories include the study of \CPV\ in 
\B-meson decay, and to over-constrain the position of
the apex of the Unitarity Triangle $(\rho, \eta)$. 
The final states of interest for the study of $\alpha$ are $B \to hh^\prime$, where $h = \rho,\pi$.
Figure~\ref{fig:feynman_diagrams} shows the Feynman diagrams
corresponding to the dominant amplitudes contributing to the
\hh\ decays.

%
%
Interference between the amplitude for direct decay, and decay
after \Bz-\Bzb\ mixing to a \hh\ final state, results
in a time-dependent decay-rate asymmetry between \Bz\ and \Bzb\
decays that is sensitive to the CKM angle $\alpha$. 
To study the time-dependent asymmetry with neutral \B-mesons, one
needs to measure the proper time difference \deltat\ between the
decay of the two \B-mesons in the event.  On reconstructing the \hh\ final
state of interest (\Brec), one needs to determine the flavor of
the other \B-meson (\Btag)\footnote{i.e. determine if \Btag\ is a
\Bz\ or \Bzb.} from its decay products. The time difference
between the decays of the two neutral \B-mesons in the event
(\Brec, \Btag) is calculated from the measured separation \deltaz\
between the $B_{\rm rec}$ and $B_{\rm tag}$ decay vertices. The
$B_{\rm rec}$ decay vertex is determined from the two charged-pion
tracks in its final state. The $B_{\rm tag}$ decay vertex is
obtained by fitting the other tracks in the event.

The signal decay-rate distribution of a \CP-eigenstate decay, $f_+
(f_-)$ for $B_{\rm tag}$= \Bz (\Bzb), is given by:
\begin{eqnarray}
f_{\pm}(\deltat) = \frac{e^{-\left|\deltat\right|/\tau}}{4\tau} [1
\pm S\sin(\deltamd\deltat) \mp \C\cos(\deltamd\deltat)]\,,
\label{eq:deltatdistribution}
\end{eqnarray}
where $\tau=1.536\pm 0.014\ps$ is the mean \Bz\ lifetime and
$\deltamd=0.502\pm 0.007 \ps^{-1}$ is the \Bz-\Bzb\ mixing
frequency~\cite{pdg}. This assumes that there is no difference
between \Bz\ lifetimes, $\Delta \Gamma = 0$. The parameters \S\
and \C\ are defined as:
\begin{eqnarray}
\S = \frac{ 2 Im \lambda     }{ 1 + |\lambda|^2}, \hspace{2.0cm}
\C = \frac{ 1 - |\lambda|^2 }{ 1 + |\lambda|^2},
\end{eqnarray}
where $\lambda=\frac{q}{p}\frac{\overline{A}}{A}$ is related to
the level of \Bz-\Bzb\ mixing ($q/p$), and the ratio of amplitudes
of the decay of a \Bzb\ or \Bz\ to the final state under study
($\overline{A}/A$). \CPV\ is probed by studying the time-dependent
decay-rate asymmetry
\begin{equation}
 {\cal A} = \frac{
  R (\deltat) - \overline{R}(\deltat) } { R (\deltat) + \overline{R}(\deltat) },
\end{equation}
where $R$($\overline{R}$) is the decay-rate for \Bz(\Bzb) tagged
events.  This asymmetry has the form
\begin{equation}
{\calA}=S\sin(\deltamd\deltat)-\C\cos(\deltamd\deltat).
\end{equation}
Belle use a different convention to \babar\ with $\C=-\acp$. 

In this article, all results are quoted using the \S\ and \C\ convention. In the
case of charged \B-meson decays (and $\pi^0\pi^0$ as there is no
vertex information) one can study a time integrated charge
asymmetry
\begin{equation}
\acp=\frac{\overline{N}-N}{\overline{N}+N},
\end{equation}
where $N$ ($\overline{N}$) is the number of \B\ ($\overline {\B}$)
decays to the final state. A non-zero measurement of \S, \C\ or
\acp\ for any of the decays under study would be a clear
indication of \CPV.

In the absence of loop (penguin) contributions in \Bz\ decays to
the \CP\ eigenstate $h^+h^-$, $\S = \sin 2 \alpha$, and $\C = 0$.
The presence of penguin contributions with different weak phases
to the tree level amplitude shift the experimentally measurable
parameter $\alpha_{\mathrm{eff}}$ away from the value of $\alpha$.
In the presence of penguin contributions $\alphaeff = \alpha +
\deltaalpha$, $\S = \sin 2 \alphaeff$, and \C\ can be non zero.

For \Bz\ decays to $\rho^+\rho^-$, $S$= \slong\ or \stran\ and $C$=
\clong\ or \ctran\ are the \CP\ asymmetry parameters for the
longitudinal and transversely polarized signal, respectively. The
decay-rate distribution for \Bz\ decays to the $\rho^\pm\pi^\mp$
final state is more complicated than the above description as
$\rho^\pm\pi^\mp$ is not a \CP\ eigenstate (discussed below).

There are two classes of measurement that the \B-factories are
pursuing. The main goal is to use \su{2}\ isospin relations to
relate different $hh^\prime$ final states and limit \deltaalpha\ in each
of the modes $\Bztopippim$, $\Bz \to \rho^\pm \pi^\mp$ and
$\Bztorhoprhom$. In some cases only weak constraints on $\alpha$
are obtained when using \su{2}\ relations and one must use a model
dependent approach to obtain a significant result.

The remainder of this article describes the isospin analysis used
to constrain \deltaalpha, and experimental techniques. There is a 
discussion of the results at the end of the article. Throughout this 
article, the experimental results are quoted
with statistical errors preceding systematic errors unless
otherwise stated.

\section{Isospin Analysis of $B\to hh^\prime$}

\begin{figure}[!h]
\begin{center}
  \resizebox{8cm}{!}{\includegraphics{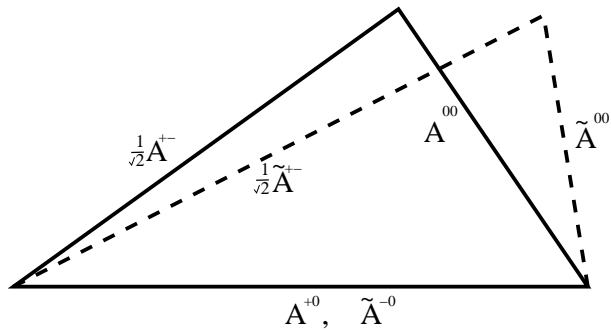}}
\end{center}
 \caption{The isospin triangle for $\B\to\pi\pi$ decays.}
\label{fig:ispintriangle}
\end{figure}

One can use \su{2}\ isospin to relate the amplitudes of \B\ decays
to $\pi\pi$ final states~\cite{gronaulondon}.  This results in two
relations: 
\begin{eqnarray}
\frac{1}{\sqrt{2}}A^{+-}=A^{+0}-A^{00},\\
\frac{1}{\sqrt{2}}\overline{A}^{+-}=\overline{A}^{-0}-\overline{A}^{00},
\end{eqnarray}
where $A^{ij}$ ($\overline{A}^{ij}$) are the amplitudes of \B\
($\overline{B}$) decays to the final state with charge $ij$. These
two relations correspond to triangles in a complex plane as shown in 
Figure~\ref{fig:ispintriangle}.

This approach only considers tree and gluonic penguin
contributions. Possible contributions from electroweak penguins
(EWP) are ignored as they do not obey \su{2}\ isospin symmetry.
EWPs have the same topology as the gluonic penguin diagram in
Figure~\ref{fig:feynman_diagrams}, with the gluon replaced by
$\gamma$ or $Z^0$ bosons. In the absence of EWP contributions
$|A^{+0}|=|\overline{A}^{+0}|$, i.e. $\acp = 0$ for $\B^+ \to
\pi^+\pi^0$. After aligning $A^{+0}$ and $\overline{A}^{+0}$, the
phase difference between $A^{+-}$ and $\overline{A}^{+-}$ is
$2\deltaalpha$.  In order to measure $\alpha$ one must measure the
branching fractions (\br) and charge asymmetries (\acp, or \C) of
\B\ decays to $\pi^+\pi^-$, $\pi^\pm\pi^0$, $\pi^0\pi^0$.

The decays $B\to\rho\rho$ are more complicated; one has an isospin
triangle relation for each of the three amplitudes in the final
state, one corresponding to the \CP\ even longitudinal
polarization and two corresponding to the \CP\ even, and \CP\ odd
parts of the transverse polarization.  The $\rho^\pm\pi^\mp$ final
state is not a \CP\ eigenstate.  This results in the need for a
pentagon isospin analysis~\cite{lipkin} or a Dalitz Plot (DP)
analysis of the $\pi^+\pi^-\pi^0$ final state~\cite{snyderquinn}
to measure $\alpha$.

There are several assumptions implicitly used in the isospin-based
direct measurements of $\alpha$.  The assumptions common to all
decay modes are (i) to neglect EWP contributions and (ii) to neglect
other \su{2}\ symmetry breaking effects.  In addition to this,
possible I=1 amplitudes~\cite{falk} in \Bztorhoprhom\ decays are
ignored.  Several groups have estimated the correction due to the
\su{2}\ breaking effect of EWP contributions to be ${\cal O} (1.5
-2)^\circ$~\cite{electroweak_pengin_calculation,zupanewp}. These
estimates consider contributions from the two EWP operators
assumed to be dominant in the effective Hamiltonian. Some
estimates of other expected \su{2}\ symmetry breaking effects
exist, where any correction is estimated to be much less than 
the current experimental precision~\cite{isospin_symmetry_breaking}.

\section{Experimental Techniques}

%
%
Continuum $\epem \to \qqbar$ ($q = u,d,s,c$) events are the
dominant background to $B\to hh^\prime$ decays. Signal \Brec\ candidates
are identified using two kinematic variables: the difference
 between the energy of the \B\ candidate and the beam
energy $\sqrt{s}/2$ in the center of mass (CM) frame \deltae; and the
beam-energy substituted mass $\mes = \sqrt{(s/2 + {\mathbf
{p}}_i\cdot {\mathbf {p}}_B)^2/E_i^2- {\mathbf {p}}_B^2}$. As the 
\B-factories collect data at the $\Upsilon(4 \rm S)$ resonance $\sqrt{s} = 10.58$ \gev.
The \B\ momentum ${\mathbf {p}_B}$ and four-momentum of the initial state
$(E_i, {\mathbf {p}_i})$ are defined in the laboratory frame.
Event shape variables are used to further discriminate between
signal and continuum background~\cite{eventshape}.

%
%
The flavor of $B_{\rm tag}$ is determined from its final state
particles.  Both \babar\ and Belle characterize the final state
particles of $B_{\rm tag}$ into events where there are leptons,
kaons and $\pi$-mesons.  The charge of these reconstructed tracks
in the final state indicate if $B_{\rm tag}$ is a \Bz\ or a \Bzb.
The signal purity varies depending on the momentum and type of
tracks in the final state. Events with leptons in the final state
are the cleanest and those with charged kaons in the final state
have less background than those with $\pi$-mesons.  In addition to
this, the probability of assigning the wrong flavor to $B_{\rm
tag}$ increases as the background increases.  \babar\ separates
events into mutually exclusive categories of events, whereas the
Belle combines this information into a single variable $r$, and
tunes event selection in bins of $r$.  The two approaches are
equivalent and are described in more detail in the following
references~\cite{tagging,babarresoandtagging}.

The \deltat\ distributions of
Equations~\ref{eq:deltatdistribution}
and~\ref{eq:deltatdistributionrp} are convoluted with a detector
resolution description, which differs for signal and continuum
background~\cite{babarresoandtagging,resolutionfunctions}, and
also take into account dilution from incorrectly assigning the
flavor of \Btag.
The signal and background parameters are extracted using an
extended unbinned maximum-likelihood fit to the data for the
analyses described here.

\section{Results}

\subsection{$B \to \pi\pi$}

The simplest decays to study in the pursuit of $\alpha$ are $B \to
\pi \pi$.  Both experiments have measured $B \to \pi^+\pi^-$,
$\pi^+\pi^0$ and $\pi^0\pi^0$ decays~\cite{babar_pipi,belle_pipi}.
The results are summarized in the Table below.  The \babar\
measurements use a data sample of 210\ifb, the Belle $\pi^+\pi^0$
result use 78\ifb, and the other Belle results use 253\ifb\ of
integrated luminosity.
All of these modes are now well established experimentally and so
it is possible to perform an isospin analysis.  The charge
asymmetry measurement of $\Bz \to \pi^0\pi^0$ accounts for the
effect of \Bz-\Bzb\ mixing.

\begin{table}[!h]
\begin{center}
\begin{tabular}{c|cccc}
mode           & Expt.   & \br\ (\e{-6})               & \S                        & \C \\
\hline
$\pi^+\pi^-$   & \babar & $5.5\pm 0.4\pm 0.3$ & $-0.30 \pm 0.17 \pm 0.03$ & $-0.09 \pm 0.15 \pm 0.04$\\
               & Belle  & $4.4\pm 0.6\pm 0.3$ & $-0.67 \pm 0.16 \pm 0.06$ & $-0.56 \pm 0.12 \pm 0.06$\\ \hline
\end{tabular}
\end{center}

\begin{center}
\begin{tabular}{c|cccc}
mode           & Expt.   & \br\ (\e{-6})               & \acp\\
\hline
$\pi^\pm\pi^0$ & \babar & $5.8\pm 0.6 \pm 0.4$        &  $-0.01 \pm 0.10 \pm 0.02$  \\
               & Belle  & $5.0 \pm 1.2 \pm 0.5$       &  $-0.02 \pm 0.10 \pm 0.01$ \\
$\pi^0\pi^0$   & \babar & $1.17\pm 0.32 \pm 0.10$     &  $-0.12 \pm 0.56 \pm 0.06$ \\
               & Belle  & $2.3^{+0.4}_{-0.5}\,^{+0.2}_{-0.3}$  &  $-0.44 ^{+0.53}_{-0.52} \pm 0.17$\\
\end{tabular}
\end{center}
\end{table}

One should note that the Belle measurement of \Bztopippim\
constitutes an observation of \CPV\ in this decay at a level of
$5.4\sigma$, and evidence for direct \CPV\ at a level of
$4.0\sigma$.  This is the second observation of \CPV\ in \B-meson
decays, the first being the measurement of a non-zero value for 
$\sin 2\beta$ from $b\to c\overline{c}s$ decays~\cite{sin2beta}. 
The \babar\ data are consistent with the hypothesis of no
\CPV\ in this decay. In recent years the two sets of results have
started to converge to a common value, but more statistics are
required to resolve the controversy.

The $\pi\pi$ isospin analysis is limited by the value of
$\br(\Bztopizpiz)$.  Unfortunately this branching fraction is
neither large enough to provide sufficient statistics with the
current data set to perform a precision measurement of
\deltaalpha, nor small enough to enable a strong bound on this
quantity.  \babar\ have performed an isospin analysis resulting in
$|\deltaalpha|<35^\circ$ (90\% C.L.).  Belle's data exclude values
of $\alpha$ between 19 and $72^\circ$ (95.4\% C.L.), and constrain
$|\deltaalpha|<38^\circ$ (95.4\% C.L.)~\cite{hazumi}.  One requires a
significant increase in statistics to perform a precision
measurement of $\alpha$ using $\B \to \pi\pi$ decays. However, it is
possible to extract a model dependent constraint on $\alpha$ from
the current results assuming \su{3}\ symmetry (i.e. exchanging $u$ and
$s$ quarks), Gronau and Rosner obtain a value of $\alpha =
(107 \pm 13)^\circ$~\cite{gronaurosner}.

\subsection{$B \to \rho\rho$}

The decay $B \to \rho\rho$ is that of a spin zero particle
decaying into two spin one particles (as shown in Figure~\ref{fig:btovv}).  As a result, the \CP\
analysis of \B\ decays to $\rho^+\rho^-$ is complicated by the
presence of three helicity states ($H=0,\pm 1$). The $H=0$ state
corresponds to longitudinal polarization and is \CP-even, while
neither the $H=+1$ nor the $H=-1$ state is an eigenstate of \CP.
The longitudinal polarization fraction $f_L$ is defined as the
fraction of the helicity zero state in the decay. The angular
distribution is
\begin{eqnarray}
&&\frac{d^2\Gamma}{\Gamma d\cos\theta_1 d\cos\theta_2}=
\frac{9}{4}\left(f_L \cos^2\theta_1 \cos^2\theta_2 + \frac{1}{4}(1-f_L) \sin^2\theta_1 \sin^2\theta_2 \right)
\end{eqnarray}
where $\theta_{i}$  $(i=1,2)$ is defined for each $\rho$-meson as the
angle between the \piz\ momentum in the $\rho$ rest frame and the
flight direction of the \Bz\ in this frame. The angle $\phi$ between the
$\rho$-decay planes is integrated over to simplify the analysis. A
full angular analysis of the decays is needed in order to separate
the definite \CP\ contributions of the transverse polarization; if
however a single \CP\ channel dominates the decay (which has been
experimentally verified), this is not necessary~\cite{Dunietz}.

\begin{figure}[!h]
\begin{center}
  \resizebox{10.0cm}{!}{\includegraphics{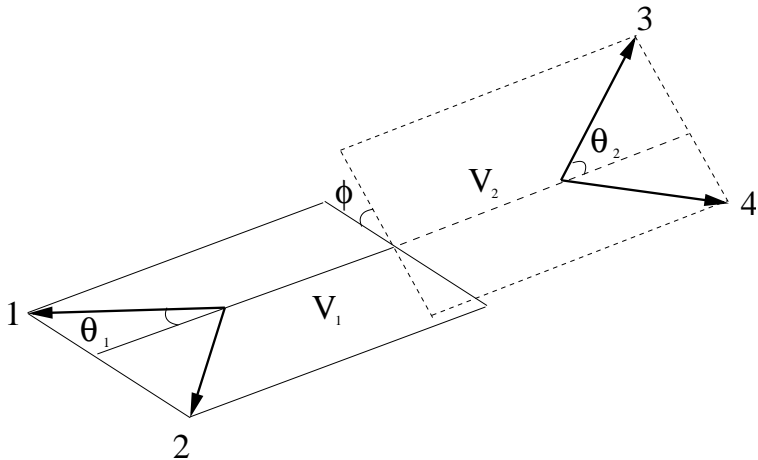}}
\end{center}
 \caption{A schematic diagram of the decay of a \B-meson via two vector particles, $V_1$ and $V_2$, into a four-particle final state.}
\label{fig:btovv}
\end{figure}

The longitudinal polarization dominates this
decay~\cite{Suzuki,my_rhorhoprl}. Not all of the $\rho\rho$ final
states have been observed, however as $\br(\Bz \to \rho^0\rho^0)$
is small, one can conservatively assume that it is longitudinally
polarized when performing an isospin analysis. The measured
branching fractions, \ptrue, and \C\ for $\B \to \rho\rho$
are summarized in the Table below, 
and $\S_L = -0.33 \pm 0.24^{+0.08}_{-0.14}$ ($0.09\pm 0.42 \pm
0.08$) for \babar. (Belle).  The \babar\ (Belle) $\rho^+\rho^-$
results use 210 (253)\ifb\ of integrated luminosity,
respectively\footnote{The branching fraction reported by \babar\
is from 82\ifb.}.  The \babar\ (Belle) $\rho^+\rho^0$ results use
82 (78)\ifb, and the upper limit on $\rho^0\rho^0$ uses 210\ifb\
of integrated luminosity.
\begin{table}[!h]
\begin{center}
\begin{tabular}{c|cccc}
mode & Expt. & \br\ (\e{-6}) & \ptrue & $\C_L$ \\ \hline
$\rho^+\rho^-$  & \babar & $30 \pm 5 \pm 4$  & $0.978\pm 0.014 ^{+0.021}_{-0.029}$  & $-0.03 \pm 0.18 \pm 0.09$  \\
                & Belle  & $24.4 \pm 2.2 ^{+3.8}_{-4.1}$ & $0.951^{+0.033}_{-0.039} \,^{+0.029}_{-0.031}$ & $0.00\pm 0.30^{+0.10}_{-0.09}$\\
\end{tabular}

\begin{tabular}{c|cccc}
mode & Expt. & \br\ (\e{-6}) & \ptrue & \acp \\ \hline
$\rho^\pm\rho^0$ & \babar & $22.5^{+5.7}_{-5.4}\pm 5.8$  & $0.97^{+0.03}_{-0.07}\pm 0.04$  & $-0.19 \pm 0.23 \pm 0.03$  \\
                 & Belle  & $31.7 \pm 7.1^{+3.8}_{-6.7}$ & $0.95 \pm 0.11 \pm 0.02$  & $0.00 \pm 0.22 \pm 0.03$   \\ \hline
$\rho^0\rho^0$   & \babar & $<1.1$ (90\% C.L.)           & -                     & -                          \\
\end{tabular}
\end{center}
\end{table}
Recent measurements of the $\Bp \to \rho^+\rho^0$ branching
fraction and upper limit for $\Bz \to \rho^0 \rho^0$~\cite{recentrhorho}
indicate small penguin contributions in $\B
\to \rho \rho$, as predicted by some calculations~\cite{aleksan}.
 The ultimate uncertainty on
$\alpha$ from $\B \to \rho\rho$ will depend on the branching
fraction and \CP\ content of $\Bz \to \rho^0 \rho^0$. The decay
$\Bz \to \rho^0 \rho^0$ has an all charged track final state and
it is possible to measure both $S$ and $C$ for the different
\CP-eigenstate components of the decay, unlike $\Bz \to \piz \piz$
decays where one can only measure $C$.

Given that penguin pollution is small, it is possible to perform
an isospin analysis of the longitudinal polarization of the
$B\to\rho\rho$ decays, and use the results of \babar's
time-dependent \CP\ analysis of $\rho^+\rho^-$~\cite{my_rhorhoprl}
to constrain $\alpha$.  If one does this, using the aforementioned
assumptions, one obtains $\alpha=(100\pm 13)^\circ$ using \babar\
data~\cite{babarrhorhoprl05}.  The error on $\alpha$ is dominated
by $|\deltaalpha| < 11^\circ$ (68\% C.L.). Belle have recently
produced a preliminary measurement of the polarization, branching
fraction and \CP\ parameters of $\Bz \to
\rho^+\rho^-$~\cite{bellerhorhoprelim}.  The Belle constraint is
$\alpha=(87 \pm 17)^\circ$, which is slightly weaker than the
\babar\ result.  Both $\rho\rho$ isospin analyses use the same
experimental information for the decays \Bztorhozrhoz, and
\Bztorhoprhoz. Figure~\ref{fig:bevan_rhorho_isospin} shows the (1-C.L.)
 plot of $\alpha$ corresponding to the isospin
analysis of the longitudinally polarized $\rho\rho$
data~\cite{ckmfitter}.  There is also a non-SM solution for $\alpha$ 
near $175^\circ$.

Our knowledge of \deltaalpha\ is primarily determined from the
experimental knowledge of the $\rho^0\rho^0$ and $\rho^\pm\rho^0$
branching ratios. As in the case of $\B \to \pi\pi$, our knowledge
on $\alpha$ is limited by the penguin contribution.  However, for
$\B \to \rho\rho$ decays \deltaalpha\ is sufficiently small to
allow the \B-factories to perform a meaningful measurement.

\begin{figure}[!h]
\begin{center}
   \resizebox{10cm}{!}{\includegraphics{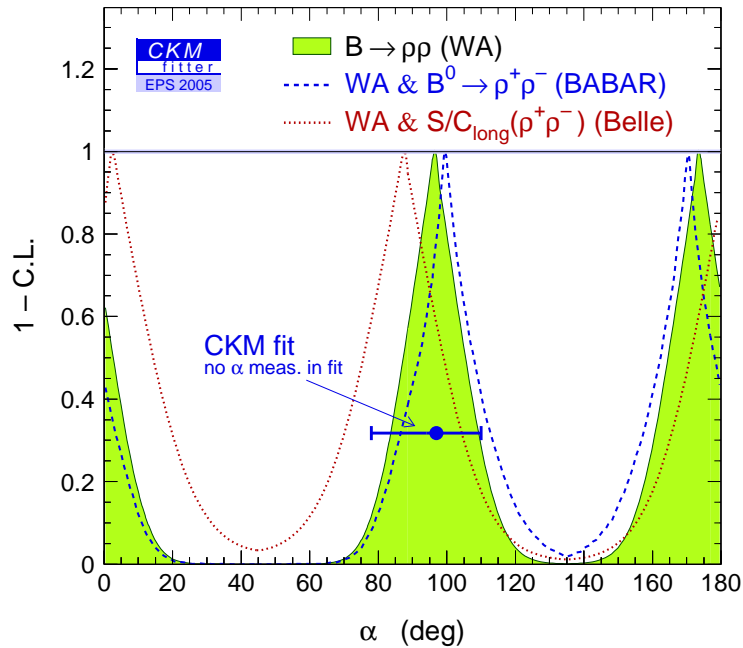}}
\end{center}
 \caption{A plot of (1 - C.L.) of $\alpha$ determined from 
          the \babar\ (dashed) and Belle (dotted) $\rho\rho$ isospin 
          analyses~\cite{ckmfitter}.  The shaded region corresponds 
          to the constraint obtained from the average of the two 
          experiments. The point with error bars corresponds to the $1\sigma$ 
          interval (68 \% C.L.) obtained from the CKM fit when excluding 
          direct measurements of $\alpha$. } \label{fig:bevan_rhorho_isospin}
\end{figure}

The $\rho\rho$ isospin triangle does not close when taking into
account the measurements of the \B-factories.  Both the CKM Fitter
and UTfit groups have noted that \su{2}\ symmetry constraints
prefer a smaller branching fraction for the $\B^+ \to
\rho^+\rho^0$ decay, at the level of $1.5-2\sigma$. The
measurements did not test for possible non-resonant backgrounds,
S-wave $\pi\pi$ contributions under the $\rho^0$, or account for
exclusive \B\ backgrounds to decays excluding charm particles. It
is important to see that this decay mode is studied in more detail
 with higher statistics in the future.

\subsection{$B \to \rho\pi$}

The decays $B \to \rho\pi$ can be analysed in two different ways.
The more straightforward approach is to cut away interference
regions of the $\pi^+\pi^-\pi^0$ DP and analyze the regions in
the vicinity of the $\rho$ resonances.  This is the so-called
Quasi-2-body approach (Q2B) and it avoids the need to understand
the interference regions. The drawbacks of the Q2B method are that
one looses a lot of statistical power by cutting on the DP, and
that interference is unaccounted for. A corollary of this is
that one requires more statistics than the \B-factories currently
have in order to obtain a significant constraint from the pentagon
isospin analysis of $B \to \rho \pi$. The alternative is to
perform an analysis of the $B\to \pi^+\pi^-\pi^0$ DP, accounting
for the interference between intersecting $\rho$ resonance bands
and other resonant structures. The Q2B approach has been studied by
\babar\ and Belle~\cite{oldrhopi,bellerhopi}, and the first
time-dependent DP analysis has been performed by
\babar~\cite{rhopidp}.

In the Q2B approach, one fits a time-dependence of
\begin{eqnarray}
f_{\pm}(\deltat) = (1\pm
\acp)\frac{e^{-\left|\deltat\right|/\tau}}{4\tau} [ (S \pm \Delta
S \sin(\deltamd\deltat) - (C \pm \Delta C)
\cos(\deltamd\deltat)]\,, \nonumber\\ \,
\label{eq:deltatdistributionrp}
\end{eqnarray}
where there are three additional parameters in comparison with Equation~\ref{eq:deltatdistribution}.  
These parameters are a charged asymmetry, \acp, between decays to $\rho^+\pi^-$ 
and $\rho^-\pi^+$ final states, and two dilution parameters; $\Delta S$ and 
$\Delta C$.
%
%
 The details of the DP analysis can be
found in~\cite{rhopidp}, where one varies a larger number of
parameters in the nominal fit, and converts these to the same
observables as the Q2B approach.
The Table below 
summarizes the experimental constraints on $\B\to\rho\pi$ (DP
analysis from \babar\ and Q2B analysis from Belle). The branching
fractions measured for this decay are $(22.6 \pm 1.8 \pm
2.2)\e{-6}$ and $(20.8^{+6.0}_{-6.3} \,^{+2.8}_{-3.1})\e{-6}$ by
\babar\ and Belle, respectively~\cite{rhopiBFs}.  These results
used 82 (29.4)\ifb\ of integrated luminosity and the time
dependent results use 192 (140)\ifb, respectively.

\begin{table}[!h]
\begin{center}
\begin{tabular}{c|ccc}
 Expt. & \S & \C & \acp \\ \hline

\babar & $-0.10\pm 0.14 \pm 0.04$ &   $0.34\pm 0.11\pm 0.05$ & $-0.088\pm 0.049 \pm 0.013$ \\
Belle & $-0.28 \pm 0.23 ^{+0.10}_{-0.08}$ &    $0.25 \pm 0.17^{+0.02}_{-0.06}$ & $-0.16 \pm 0.10 \pm 0.02$\\
\end{tabular}
\end{center}
\end{table}

Using \babar's DP result, one obtains the following constraint;
$\alpha=(113^{+27}_{-17} \pm 6)^\circ$. Figure~\ref{fig:bevan_rhopi_isospin} shows the corresponding 
(1 - C.L.) plot for $\alpha$. This result is self-consistent as the
strong phase differences and amplitudes are determined solely from
the structure of the DP. The unique aspect of this result is that
there is only a single solution between 0 and $180^\circ$, and thus a
two fold ambiguity on $\alpha$. As a result this measurement is an
important constraint on $\alpha$, ancillary to that from $\B
\to\rho\rho$.

\begin{figure}[!h]
\begin{center}
  \resizebox{10cm}{!}{\includegraphics{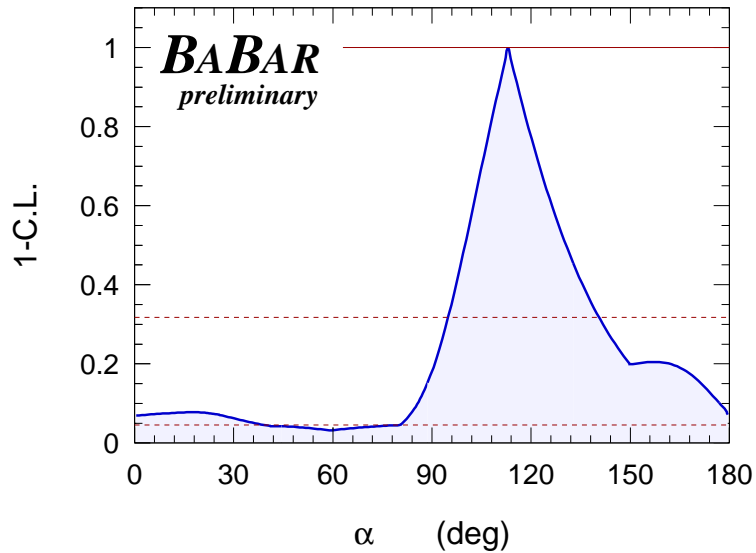}}
\end{center}
 \caption{A plot of (1 - C.L.) of $\alpha$ determined from
          \babar's $\pi^+\pi^-\pi^0$ DP analysis~\cite{rhopidp}.}

\label{fig:bevan_rhopi_isospin}
\end{figure}

It is possible to derive a model dependent constraint on $\alpha$
from these data as shown by Gronau and Zupan~\cite{gronauzupan}.
One can make several assumptions, including (i) the relative
strong phase differences between tree level contributions for
$\rho^+\pi^-$ and $\rho^-\pi^+$ being less than $90^\circ$ (ii)
\su{3} being exact for penguin amplitudes (iii) neglecting EWP
(iv) neglecting annihilation and exchange diagrams to derive a
constraint on $\alpha$. \su{3} breaking effects are estimated in
this model using CLEO, \babar\ and Belle data. When doing this
Belle obtains $\alpha=(102\pm 11_{expt} \pm 15_{model})^\circ$.
Gronau and Zupan compute the corresponding result for the \babar\
analysis as $(93\pm 4_{expt} \pm 15_{model})^\circ$.  There are
other models proposed in the literature, for
example~\cite{gronaulunghi}.

\section{Discussion}

The indirect measurements of $\alpha$ from the SM are $(98.2 \pm
7.7)^\circ$~\cite{utfitter} and
$(97^{+13}_{-19})^\circ$~\cite{ckmfitter} from the UTfit and CKM
Fitter groups, respectively. Direct measurements from the
\B-factories are in good agreement with these predictions, where
the uncertainty on $\alpha$ from $\B \to hh^\prime$ decays is known to
${\cal O}(9^\circ)$.  The precision of this result is dominated by
an isospin analysis of $\B \to \rho\rho$ decays, with an important
contribution from $\B\to\rho\pi$ as the latter result suppresses
a second solution for $\alpha$ near $175^\circ$. This solution is 
disfavored at the $1.3\sigma$ level by studies of 
$\Bz\to\pi^+\pi^-\piz$ decays.  The \B-factory constraints
on $\alpha$ will be improved as additional data are analyzed over
the next few years, and these measurements will start to test the
closure of Unitarity Triangle with precision. A significant
isospin analysis result from $\B\to\pi\pi$ requires a considerable
increase in statistics.  The measurement of $\alpha$ from
$\B\to\pi\pi$ remains a vital part of the program, as theoretical
uncertainties in the extraction of $\alpha$ (other than the
determination of \deltaalpha) are the best understood of all the
channels that have been discussed here.

Similarly, the model dependent calculations on $\alpha$ are in
agreement with the SM.  The constraints from $\B\to \pi\pi$ and
$\rho\pi$ decays have a precision of $13^\circ$ and $15.5^\circ$,
respectively.  These constraints are dominated by theoretical
uncertainties.  The precision on the model dependent constraints
obtained are comparable with the isospin analysis result from $\B
\to \rho\rho$.

The measurement of $\alpha$ using $\B \to a_1 \pi$ decays was
proposed some time ago~\cite{a1pialeksan}. The recent observation
of $\Bz \to a_1^+ \pi^-$ with a branching fraction ${\cal O}( 45
\e{-6})$~\cite{a1pi} raises the question of how well one can
ultimately measure $\alpha$ using this decay mode.  The use of
\su{3}\ to measure $\alpha$ with $\B \to a_1 \pi$ decays has
recently been proposed~\cite{gronauzupan2005}.

Studies of the LHCb experiment's~\cite{lhcb} sensitivity to $\alpha$ using
$\B \to \pi^+\pi^-\piz$ decays have been
performed~\cite{deschamps}. LHCb is expected to achieve a  $10^\circ$ precision 
on $\alpha$ within the first year of data taking. The physics reach of decays involving
$b \to u$ transitions with all charged particles in the final
states, such as $\Bz \to \rho^0\rho^0$ and $\Bz \to \aone^+
\pi^-$ should also be investigated. 
The current \B-factories will 
reach a precision of $7-10^\circ$ on $\alpha$ with 1 \iab\ of 
data using $\B\to\rho\rho$ decays.  The ultimate precision 
reached depends on the values of the branching fraction, \ptrue, $\S$ and $\C$ in 
$\Bz \to \rho^0\rho^0$.  Sensitivity projections for the $\pi^+\pi^-\pi^0$ DP analysis
are very dependent on the many input parameters involved.  For this reason, the running 
experiments have not extrapolated the precision of $\alpha$ obtained from 
$\pi^+\pi^-\pi^0$ decays to higher luminosities.

\section*{Acknowledgments}

This work is supported in part by the UK Particle Physics and
Astronomy Research Council (PPARC), and the U.S. Department of
Energy under contract number DE-AC02-76SF00515.  The author wishes
to thank K. K. Phua for the opportunity to write this review, and
I. Bigi, R. Faccini and J. Zupan for useful discussions
and comments during the preparation of this article.

\end{document}